\documentclass[%
 reprint,
 amsmath,amssymb,
]{revtex4-1}

\usepackage{braket}
\usepackage{courier}
\usepackage{graphicx}
\usepackage{dcolumn}
\usepackage{bm}

\begin{document}

\preprint{APS/123-QED}

\title{Unsupervised Learning Eigenstate Phases of Matter}

\author{Steven Durr}
\author{Sudip Chakravarty}%
\affiliation{%
Mani L Bhaumik Institute for Theoretical Physics\\
Department of Physics and Astronomy, University of California Los Angeles, Los Angeles, California 90095, USA
}%


\date{\today}

\begin{abstract}
Supervised Learning has been successfully used to produce phase diagrams and identify phase boundaries when local order parameters are unavailable. Here, we apply unsupervised learning to this task. By using readily available clustering algorithms, we are able to extract the distinct eigenstate phases of matter within the transverse-field Ising model in the presence of interactions and disorder. We compare our results to those found through supervised learning and observe remarkable agreement. However, as opposed to the supervised procedure, our method requires no strict assumptions concerning the number of phases present, no labeled training data, and no prior knowledge of the phase diagram. We conclude with a discussion of clustering and its limits.

\end{abstract}

\maketitle


\section{\label{sec:into}Introduction}
Recently, machine learning has been applied to the task of identifying phases of matter -- particularly in scenarios in which local order parameters are not available~\cite{Venderley2018MachineLO, Jnsson2018DetectingTM, 1605.01735, Schindler2017ProbingML, 1703.02435, PhysRevB.99.085406, 1805.05961, PhysRevE.99.023304, 1808.00084, 1606.00318, PhysRevB.96.184410, 1707.00663, 1608.07848, PhysRevLett.118.216401, Carleo602, 1706.08111, 1705.01947, 1612.04909, 1610.02048, PhysRevE.95.062122, PhysRevB.97.134109}. Approaches have largely focused on the application of supervised learning. Using this technique, data is sampled from points in parameter space known to belong to a certain phase. Some function (often a neural network) is then trained to predict the phase given input data. If the function is able to effectively generalize, it is then possible to apply it to data coming from points across the parameter space. This allows one to produce a phase diagram, and gain insight into the underlying physics.

Techniques from unsupervised learning have also been shown to be effective at identifying phases~\cite{1703.02435, 1805.05961, 1808.00084, 1606.00318, PhysRevB.96.184410, 1707.00663, PhysRevE.99.023304, PhysRevE.95.062122, PhysRevB.97.134109}. When applied to the 2D Ising model, for instance, tools such as autoencoders have been able to extract a local order parameter~\cite{1606.00318, PhysRevE.95.062122}. In addition, clustering techniques have been used to accurately distinguish spin data known to correspond to distinct topological sectors~\cite{1805.05961}.

Here, we expand on work applying unsupervised learning towards identifying phases. Using readily available clustering algorithms, we explore the parameter space of a system known to display many-body localization and eigenstate phase transitions. In particular, the problem we study has been effectively treated using supervised learning~\cite{Venderley2018MachineLO, Jnsson2018DetectingTM}, which, when compared to conventional techniques, was able to predict the sharpest phase boundary to date~\cite{Venderley2018MachineLO}.

We therefore use this supervised approach as a starting point from which to compare our unsupervised method, and find that we are able to produce highly similar results. Notably, our method relies on no separate training data, no prior knowledge of the phase space, and even no explicit assumption of the number of phases present. 

We conclude with a discussion of what can be learned from this success, as well as the role and usefulness of machine learning algorithms for the task of identifying phases.

\section{\label{sec:cmbl}Clustering Many-Body Localized Phases}

Generally, an isolated interacting quantum system is said to display many-body localization (MBL) if it fails to thermalize under its own unitary time evolution. On the other hand, a quantum system is said to be thermal if it is able to serve as its own heat bath. Different MBL phases exist, displaying different symmetries and topological order. The transition of a state among these thermal and MBL phases represents a dynamic eigenstate phase transition -- for which an extensive theoretical description does not currently exist.

Here we use clustering algorithms to analyze eigenstate phase transitions within the transverse-field Ising model in the presence of interactions and disorder:
$$
H = -\sum_{i=1}^{L}\big( J_{i} \sigma_{i}^{z} \sigma_{i+1}^{z} +h_{i} \sigma_{i}^{x} + \lambda (\bar{h} \sigma_{i}^{x} \sigma_{i+1}^{x} +\bar{J}\sigma_{i}^{z}\sigma_{i+2}^{z})\big)
$$

Above, $\sigma_{i}^{\alpha}$ are the Pauli matrices and $\{J_{i}\}$ and $\{h_{i}\}$ are log-normal distributions with respective means $\bar{J}$ and $\bar{h}$, and the standard deviations of their logarithms equal to $1$. We use open boundary conditions and a length $12$ spin chain. 

The limits of this model have been studied, and are known to exhibit different eigenstate phases~\cite{PhysRevB.88.014206, PhysRevX.4.011052, PhysRevLett.113.107204, PhysRevB.51.6411}. In particular, for $\overline{h} \gg \overline{L}, \lambda$, the system is expected to exhibit a many-body localized paramagnetic phase (MBL PM). In this limit, MBL PM eigenstates roughely correspond to product states in the $\sigma_{x}$ basis (e.g. $\ket{\leftarrow \leftarrow \rightarrow \cdot\cdot\cdot}$). 
In the opposite limit of  $\overline{J} \gg \overline{h}, \lambda$ the system is expected to be in a many-body localized spin-glass phase (MBL SG). Here, states resemble global superpositions of spins in the $\sigma_{z}$ basis with frozen domain walls (e.g. $\frac{1}{\sqrt{2}}(|\ket{\uparrow\downarrow\uparrow\uparrow \cdot\cdot\cdot}\pm\ket{\downarrow\uparrow\downarrow\downarrow \cdot\cdot\cdot}$).
In the limit of $\lambda \gg \overline{J} = \overline{h}$ the system is expected to be in a thermal phase. In addition, the model is self dual about $\overline{\text{Log}(J)}=\overline{\text{Log}(h)}$. Therefore, up to the effects of finite size and open boundary conditions, we should observe symmetry about this point in the phase diagram.

Supervised learning has been used to study this model~\cite{Venderley2018MachineLO, Jnsson2018DetectingTM} and determine phase boundaries with more precision. In~\cite{Venderley2018MachineLO}, the authors demonstrate that supervised learning can accurately identify phase boundaries with a high degree of clarity. Therefore, to evaluate the success of our analysis, we will compare our results to those found through the use of supervised learning following the procedure of ~\cite{Venderley2018MachineLO}.

\subsection{\label{sec:pd}Producing Data}

We vary two parameters: $\lambda$ and $\Delta_{Jh}:= \overline{\text{Log}(J)}-\overline{\text{Log}(h)}$, and obtain data for a grid of points in parameter space where $\lambda \in [0, 2]$ and $\Delta_{Jh} \in [-3, 3]$~\cite{Venderley2018MachineLO}. At each point, we obtain the Hamiltonian and find its eigenvectors \footnote{We exclude the highest and lowest 10\% of the eigenvectors to reduce potential deviation from the trend of a given phase}. For each eigenvector, $\ket{\psi}$, we can calculate the reduced density matrix as follows:

We consider the system to be split into two parts: region A containing the middle $4$ spins, and region B containing the outer $8$. We then trace over the states of B to obtain the reduced density matrix: 
$$
\rho_{A} = Tr_{\mathcal{H} \setminus \mathcal{H_{A}}}\big( \ket{\psi}\bra{\psi} \big)
$$

The $-\text{Log}$ of the eigenvalues of $\rho_{A}$ would give us the entanglement spectrum -- known to carry information concerning many body localization~\cite{1603.00880, PhysRevB.82.174411, article}. The eigenvalues of $\rho_{A}$ themselves are probabilities, giving us a vector in $2^{4} = 16$ dimensions with elements summing to $1$.

There exist established distance metrics motivated by information theory for expressing the similarity of two probability distributions. For this reason, combined with the additional benefit of having the data live in a compact region, we use the probability vectors corresponding to the eigenvalues of the reduced density matrix as our data for clustering. We take our distance metric between two probability vectors to be the Jensen-Shannon distance~\cite{1207388}, which is a bounded metric expressing similarity between probability distributions.

Therefore, for each disorder realization and each point in parameter space, we calculate an array of lists containing reduced density matrix eigenvalues (one list of eigenvalues for each eigenvector). Here we generate $100$ disorder realizations and evaluate Hamiltonians at $1200$ points in parameter space.

\subsection{\label{sec:cd}Clustering Data}

We collect $N=1000$ samples of our data, each with $n=5$ elements taken at each of the $1200$ positions in parameter space ($=6000$ elements per sampling). 

As a first step, we apply the HDBSCAN clustering algorithm~\cite{McInnes2017} (Hierarchical Density-Based Spatial Clustering of Applications with Noise) to run an exploratory analysis and identify structure within the data set. HDBSCAN is ideal for exploratory clustering due to its lack of hard assumptions about the data. In particular, it does not assume clusters to be convex, nor does it assume a set number of clusters to search for. A more complete discussion of HDBSCAN can be found in the appendix. 

We next set HDBSCAN's two main parameters. We set the minimum cluster size (\texttt{min\char`_cluster\char`_size}) by using a rough prior concerning the size of the smallest cluster we expect to see. This clearly varies based on application. Here we specify that we are interested in finding clusters which comprise at least $1/10$ of the total data set. This gives us a minimum cluster size of $6000/10 = 600$ elements. 

\begin{figure}[h]
    \centering
    \includegraphics[width=0.5\textwidth]{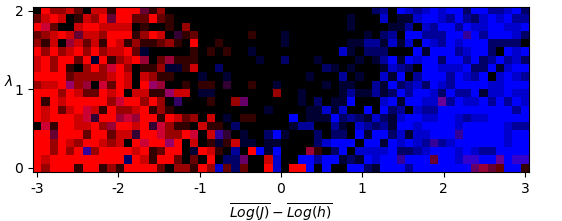}
    \caption{A phase diagram found by applying the HDBSCAN clustering algorithm to a single sampling of data from across parameter space. Two clusters are identified, one to the left (red) and one to the right (blue). A darker region in the center corresponds to data classified as low-density noise.}
    \label{fig:hdbscan_single}
\end{figure}

A suitable value for the \texttt{min\char`_samples} parameter can be set by evaluating the density-based validation score~\cite{inproceedings} across a range. The value of \texttt{min\char`_samples}, however, is observed to have little effect on the clustering. By identifying cluster labels with their phase space positions, we may then obtain a corresponding phase diagram (Fig. \ref{fig:hdbscan_single}). 

\begin{figure}[h]
    \centering
    \includegraphics[width=0.5\textwidth]{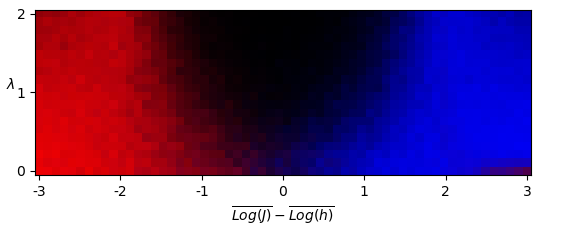}
    \caption{A phase diagram obtained by averaging $1000$ samples of data from across parameter space, each clustered using HDBSCAN clustering. Clusters from each sample of data were identified with one another using their relative $\Delta_{Jh}$ position -- the cluster with lower mean value was colored red, and the cluster with the higher value colored blue. Averaging over all diagrams gives us the above result.}
    \label{fig:hdbscan_average}
\end{figure}

By examination, we can see that there are two observed clusters which may be ordered by their average $\Delta_{Jh}$ value. We then apply HDBSCAN to all $1000$ samples, and note that in over $98\%$ of samples, two such clusters were formed -- the remaining $<2\%$ forming only a single cluster. We discard these, identify clusters by their weighted $\Delta_{Jh}$ order, and average over samples to obtain a phase diagram (Fig. \ref{fig:hdbscan_average}).

By inspection, we can identify three regions of the phase diagram: two to the left and right identified as clusters, and a third in the center considered noise. 

HDBSCAN does not attempt to allocate every element into a cluster. Rather, it assumes that data sets may contain sparse noise not belonging to any particular cluster. Here, we can see that the data HDBSCAN considers noise may instead form a third, more sparse cluster. 

\begin{figure}[h]
    \centering
    \includegraphics[width=0.5\textwidth]{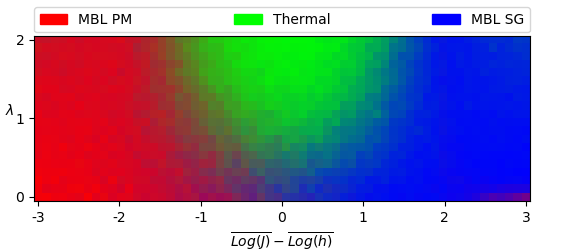}
    \caption{\textbf{Unsupervised Phase Diagram:} The phase diagram found by applying spectral clustering to data from across parameter space using a total of $100$ disorder realizations, and clustering $1000$ subsets of data sampled from these realizations. During clustering, we use reduced density matrix eigenvalues as data, and the Jensen-Shannon distance as our metric.}
    \label{fig:spectral_pd}
\end{figure}

To investigate this, we apply an algorithm known as spectral clustering~\cite{scikit-learn} (discussed in more detail in the appendix). While still not assuming convexity of clusters, spectral clustering allows us to include assumptions about the number of clusters to form within the data. We repeat the above procedure, taking the number of clusters to form to be $3$, and again using Jensen-Shannon distance. Averaging over $1000$ samples, we obtain a total phase diagram (Fig. \ref{fig:spectral_pd}).

\section{\label{sec:aor}Analysis of Results}

\begin{figure}[h]
    \centering
    \includegraphics[width=0.5\textwidth]{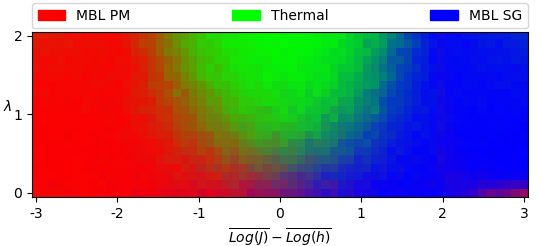}
    \caption{\textbf{Supervised Phase Diagram:} The phase diagram found by applying a trained neural network to the entanglement spectra from across parameter space using $100$ disorder realizations.}
    \label{fig:supervised_pd}
\end{figure}

We see that the sparse data considered noise by HDBSCAN was successfully identified as a third cluster. Physically, we do expect three phases to be present: two clusters corresponding to a many-body localized paramagnetic phase and a many-body localized spin-glass phase (referred to as MBL PM and MBL SG, respectively), and one cluster corresponding to a thermal phase which conforms to the eigenstate thermalization hypothesis. 

Up to the effects of finite system size and open boundary conditions, we expect our phase diagram to be symmetric under $\Delta_{Jh} \rightarrow -\Delta_{Jh}$, and for the phases to exist in the general regions identified within the phase diagram.
Therefore, we appear to have formed clusters corresponding to each of these predicted phases.

\begin{figure}[h]
    \centering
    \includegraphics[width=0.5\textwidth]{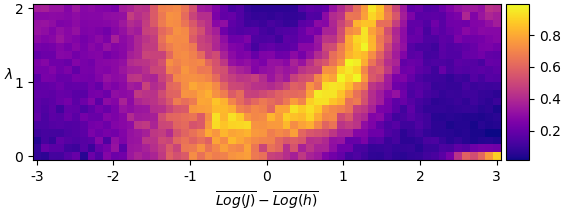}
    \caption{\textbf{Unsupervised $\mathcal{C}$ Diagram:} The measure of confusion, $\mathcal{C}$, plotted across parameter space obtained through spectral clustering.}
    \label{fig:spectral_c}
\end{figure}

\begin{figure}[h]
    \centering
    \includegraphics[width=0.5\textwidth]{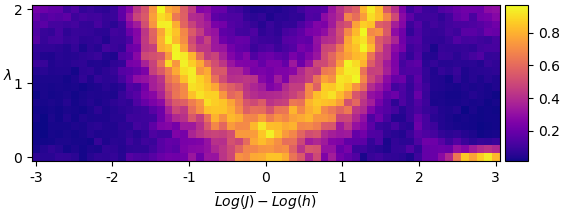}
    \caption{\textbf{Supervised $\mathcal{C}$ Diagram:} $\mathcal{C}$ plotted across parameter space obtained using a trained neural network.}
    \label{fig:supervised_c}
\end{figure}
\subsection{\label{sec:ctsl}Comparison to Supervised Learning}

Following the procedure outlined in~\cite{Venderley2018MachineLO} and using the framework of TensorFlow~\cite{tensorflow2015-whitepaper}, we produce a phase diagram for the system by applying a trained neural network to our data. We can then compare this to the phase diagram found through unsupervised learning. 

We train a neural network on a set of simulated data emanating from three points in $\Delta_{Jh}$, $\lambda$ space:
$$
(0.8, 0.2), (-0.8, 0.2), (0.0, 1.0)
$$
These correspond to the MBL SG, MBL PM, and Thermal phases, respectively. After training, we apply the neural network to data from across the parameter space and use the resulting phase predictions to produce a phase diagram (Fig. \ref{fig:supervised_pd}).

To measure uncertainty in a phase assignment, in ~\cite{Venderley2018MachineLO} the authors define a quantity $\mathcal{C}$. If each point on the final phase diagram has a corresponding probability vector, $\vec{p} = (p_{1}, p_{2}, p_{3})$, then  $\mathcal{C}$ is defined as $\mathcal{C} = 1-\overline{\text{d}}_{\text{min}}$. Here $\text{d}_{\text{min}} = \text{min}{| \vec p - \vec v  | : \vec{v} \in Q}$, and Q is the set of points of extremal phase uncertainty: 
$(1/2, 1/2, 0)$, $(1/2, 0, 1/2)$, $(0, 1/2, 1/2)$, and $(1/3, 1/3, 1/3)$. $\overline{\text{d}}_{\text{min}}$ is then the value of $\text{d}_{\text{min}}$ normalized by its maximum possible value.

We calculate this measure for both the unsupervised (Fig. \ref{fig:spectral_c}) and the supervised (Fig. \ref{fig:supervised_c}) approaches, and compare the two. 

\begin{figure}[t]
    \centering
    \includegraphics[width=0.5\textwidth]{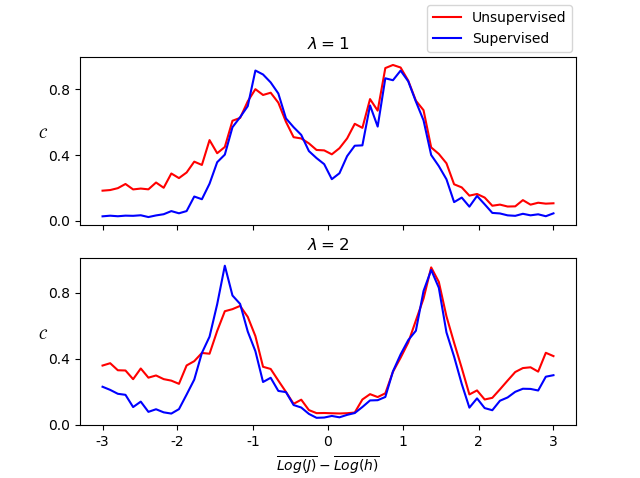}.
    \caption{Slices of the $\mathcal{C}$ diagrams taken at $\lambda = 1$ and $\lambda = 2$ showing both the supervised and unsupervised results.}
    \label{fig:c_slices}
\end{figure}

Qualitatively, the two diagrams are similar. We examine this agreement by taking slices of $\mathcal{C}$ at constant $\lambda$. Again, we find agreement at both $\lambda = 1$ and $\lambda = 2$ (Fig. \ref{fig:c_slices}). Note, however, that the supervised method produces consistently lower values of $\mathcal{C}$ away from phase transitions. 

\begin{figure}[h]
    \centering
    \includegraphics[width=0.5\textwidth]{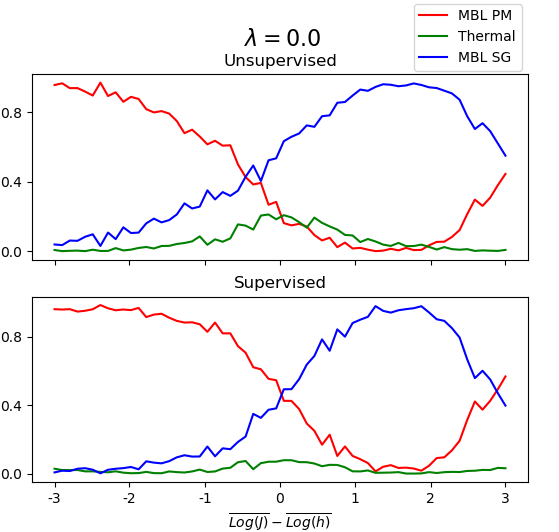}
    \caption{Slices of the unsupervised (top) and supervised (bottom) phase diagrams along $\lambda = 0$. Note the asymmetry near $\overline{\text{Log}(J)}-\overline{\text{Log}(h)} = 3$ is found in both procedures.}
    \label{fig:lambda_0}
\end{figure}

We also observe agreement at $\lambda = 0$, where both methods observe the same asymmetry at $\Delta_{Jh}$ = 3, present potentially due to open boundary conditions and finite-size effects (Fig. \ref{fig:lambda_0}). 

\begin{figure}[t]
    \centering
    \includegraphics[width=0.5\textwidth]{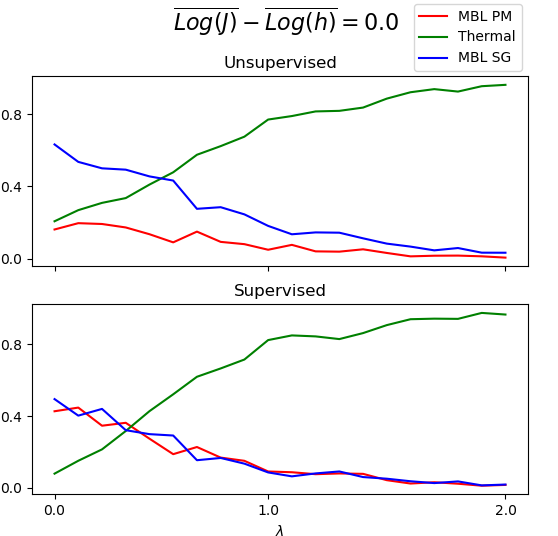}
    \caption{Slices of the phase diagrams  found using both unsupervised (top) and supervised (bottom) methods, and taken along $\overline{\text{Log}(J)}-\overline{\text{Log}(h)} = 0$. The y-axis indicates the probability of classification into a given phase. Note that both methods predict the similar presence of a thermal phase as $\lambda$ increases,}
    \label{fig:mlj_mlh_0}
\end{figure}

\section{\label{sec:disc}Discussion}
Here, we have outlined an unsupervised method to study the phase space of a system demonstrating MBL transitions. Our method applies readily available clustering algorithms to segment phase space into three regions. Specifically, we apply HDBSCAN and spectral clustering to the eigenvalues of reduced density matrices, using Jensen-Shannon distance as a metric for clustering.

When compared to the corresponding result obtained through supervised learning, we find remarkable agreement between the phase boundaries that both methods predict. Both techniques are able to produce these meaningful results using small data sets. Here we relied on a set of only $100$ disorder realizations. Our method, however, requires no strict assumptions about the number of clusters present, no labeled training data, and no prior knowledge of the phase diagram. 

\begin{figure}[h!]
    \centering
    \includegraphics[width=0.5\textwidth]{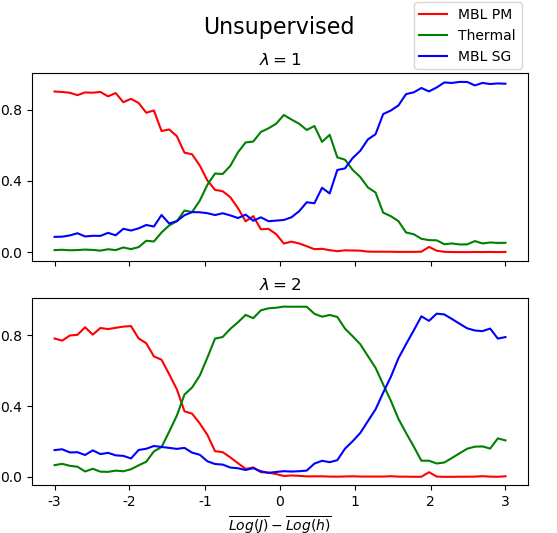}
    \caption{Slices of the unsupervised phase diagram taken at $\lambda = 1$ (top) and $\lambda = 2$ (bottom) showing the probability of classification into one of three phases.}
    \label{fig:spectral_lambda_slices}
\end{figure}

\begin{figure}[h!]
    \centering
    \includegraphics[width=0.5\textwidth]{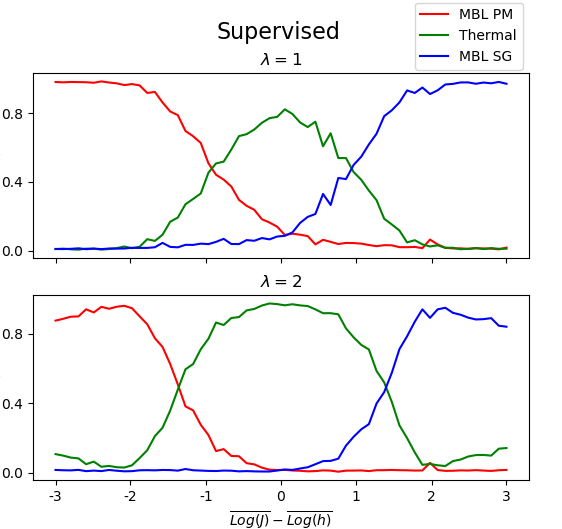}
    \caption{Slices of the supervised phase diagram taken at $\lambda =1$ (top) and $\lambda = 2$ (bottom) showing the predicted probabilities of each phase.}
    \label{fig:supervised_lambda_slices}
\end{figure}

In the case of supervised learning applied to eigenstate phases, it is not readily apparent which features are being extracted from the data that would indicate the presence of a particular phase. Therefore, this is a black box method -- an issue present in many applications of machine learning. In our use of clustering, the same problem exists. 

When applying a clustering algorithm, data is segmented into groups according to each algorithm's implicit conception of what a cluster comprises. Moreover, there does not exist -- and cannot exist~\cite{NIPS2002_2340} -- a satisfying universal axiomatic approach to define the goals of clustering. Rather, trade-offs between different clustering criteria are intrinsic. These trade-offs can be seen in practice. Clustering algorithms often fail to perform when the ideal clusters are of very different sizes, different densities, and different shapes. Algorithms which perform well in one of these situations may fail in another. 

Choosing a function to express the similarity between two elements (i.e. a distance function) also often relies on heuristics. Distance functions can be chosen based on the nature of the data at hand and the goal of the clustering. Other tools from unsupervised learning can be effective here. Autoencoders, for instance, map elements from an original space to a latent space. Spatial separation of two data points in the latent space then corresponds to some meaningful difference between the two initial elements. A set of words, for example, can be mapped to vectors in a latent space. Spatial similarity between vectors in this latent space (e.g. cosine similarity) corresponds to similarity in meanings of the words. Forming clusters in this latent space can then yield meaningful groupings. 

From the perspective of clustering, there does not necessarily exist an a priori \lq correct' partitioning of the data. The optimal clustering of data is instead dependent on the application. HDBSCAN uses a quantified heuristic expressing hierarchical depth within the data to determine the number of clusters to form. Other methods to determine the optimal number of clusters are also available, but all rely on optimizing some conception (either stated explicitly or implied) of what a cluster should be. 

With these concerns in mind, when using a clustering algorithm whose optimization criteria is without a direct mapping to physics, one cannot be immediately sure that the resulting partitioning of physical data must usefully correspond to distinct physical categories. Nonetheless, we have demonstrated that the clustering procedure applied here has yielded useful results. Our goal, however, is not to show that clustering techniques accurately and reliably extract MBL phase boundaries. Rather, it is to show that meaningful relationships within physical data can be quickly and cheaply explored by using reasonably applied clustering techniques. 

This work was supported in part by funds from
David S. Saxon Presidential Term Chair at UCLA.

%

\section{\label{sec:A}Appendix}
\subsection{\label{sec:level2}High level Overview of Relevant ML}

{\setlength{\parindent}{0cm}
\textbf{Supervised Learning}
}

Supervised learning makes use of a function, $f$, to classify data into $n$ categories. 

$$
f_{W}:R^{N} \rightarrow R^{n}
$$
with parameters $W$, as well as $n$ sets of labeled training data -- each representative of its corresponding category of data. We wish to interpret the output of $f_{W}(v)_{i}$ to obtain the probability that a data point, $v \in R^{N}$, belongs to the $i^{th}$ data set. To do this, we normalize $f_{W}$ using the softmax function:
$$\text{softmax}(v) = \frac{Exp(v)}{\sum_{i} Exp[v_{i}]}$$
And interpret
$$
P(v \in \text{Category}_{i}) = \text{softmax}(f_{W}(v))_{i}
$$

We then use the training data sets to find optimal parameters $W^{\prime}$, such that for $v \in T_{i}$, $\text{softmax}(f_{W^{\prime} }(v))_{i}$ is maximized, while for $j \neq i$, $\text{softmax}(f_{W^{\prime}}(v))_{j}$ is minimized. This is achieved through the minimization of some loss function which characterizes the error of the prediction (for example, cross entropy). 

If care is taken to avoid overfitting and each training set is sufficiently representative of its corresponding category of data, then $\text{softmax} \circ f_{W^{\prime}}$ can be interpreted as a function which probabilistically categorizes our data into $n$ sets. In particular, $f_{W}$ is often chosen to take the form of a neural network. 

\hspace{1cm}

{\setlength{\parindent}{0cm}
\textbf{Unsupervised Learning}
}

In unsupervised learning, we do not require data to be labeled. Rather, we follow a procedure to extremize some quantity in order to identify structure present in the data. Clustering is a form of unsupervised learning whose goal is to separate data into groups such that elements within a group are in some sense similar, and elements between groups are different. 

Different clustering algorithms use different approaches to group data. Below, we describe some meaningful differences in approaches that clustering algorithms can take. 

\hspace{1cm}

{\setlength{\parindent}{0cm}
\textbf{Parametric vs Density-Based}
}

Parametric clustering algorithms assume (either implicitly or explicitly) that the data takes a certain form. This could be that the clusters are convex (as assumed by algorithms such as k-means) or that the pdf from which the data points are drawn are sums of Gaussians (as assumed by a Gaussian mixture model). Furthermore, these models generally assume knowledge of the number of clusters present in advance. 

On the other hand, density-based clustering assumes that the data is generated according to a probability distribution and seeks to identify connected components of level sets of the pdf. In practice, these algorithms separate high density regions of the data from low density regions. Connected components of these high density regions are then considered clusters. 

\hspace{1cm}

{\setlength{\parindent}{0cm}
\textbf{Flat vs Hierarchical}
}

Flat clustering algorithms require us to set a parameter identifying the \lq granularity' of the clusters we would like to form. For parametric algorithms, this could correspond to specifying the number of clusters to form in the data (i.e. the resolution of clustering). For density-based clustering algorithms, this might correspond to choosing which level set of the pdf to use in clustering. Different level sets may then yield different connected components. 

Hierarchical clustering algorithms avoid setting a granularity parameter. Rather, they construct a hierarchy of groupings, with similar clusters merging into one another as we decrease the resolution. 

\hspace{1cm}

{\setlength{\parindent}{0cm}
\textbf{HDBSCAN}
}

HDBSCAN is a density-based clustering algorithm which also uses tools from hierarchical clustering. It builds upon DBSCAN, a flat density-based algorithm, and can be ideal for exploratory clustering. 

In exploring an unknown data set, we would like our clustering algorithm to make as few assumptions about the data as possible. These include assumptions about the number of clusters present, as well as the shape of those clusters.

As the HDBSCAN algorithm is density-based, it does not assume that clusters must be of a specific form. In addition, instead of assuming a specific number of clusters to find, HDBSCAN uses a hierarchical analysis of the data to quantify the \lq depth' of potential clusters. Its hierarchical technique allows HDBSCAN to predict which clusters to form, based on how resilient their presence is under variation of the clustering resolution. Furthermore, HDBSCAN does not attempt to segment each point into a cluster. Rather, it assumes that clusters may be surrounded by lower density noise. 

HDBSCAN has two parameters: \texttt{min\char`_cluster\char`_size}, and \texttt{min\char`_samples}. The \texttt{min\char`_cluster\char`_size} parameter simply puts a lower bound on the size of clusters to form. \texttt{min\char`_samples} is less intuitive. It expresses how conservative or aggressive a given clustering of the data should be. A greater value corresponds to a more conservative clustering and more points being declared as noise. Its value generally does not radically affect the final partitioning. However, some quantitative reference for a suitable value of \texttt{min\char`_samples} can be found by applying the density-based validation score to resulting clusters. 

\hspace{1cm}

{\setlength{\parindent}{0cm}
\textbf{Spectral Clustering}
}

Spectral clustering is a density-based clustering algorithm related to manifold learning and the DBSCAN algorithm. It allows the user to include specific assumptions concerning the number of clusters to form. Spectral clustering approaches separating and grouping data as a graph partitioning problem. Given a distance metric, $d$, the algorithm generates a similarity matrix. Typically this is done using a Gaussian kernel similarity function:
$$
S_{i, j} = e^{-d_{i, j}^{2} / (2 \sigma^{2})}
$$
Where $\sigma$ corresponds to the size of neighborhoods expected to form within the data.  

The algorithm then computes the Laplacian matrix of the resulting graph and collects the first $k$ eigenvectors. Data is then projected to the $k$ dimensional vector space spanned by these eigenvectors and clustered in this space using a more simple algorithm (such as k-means). This process corresponds to moving to a vector space in which position expresses connectivity, and clustering in this space.
\end{document}